\documentclass{PoS}

\usepackage{xspace}
\usepackage{textcomp}

\newcommand{\mec}{$\mu$--$e$ conversion\xspace}
\newcommand{\unit}[1]{\ensuremath{\,\textrm{#1}}\xspace}
\title{Status of the AlCap experiment}

\ShortTitle{Status of the AlCap experiment}

\author{\speaker{R. Phillip Litchfield}\thanks{On behalf of the AlCap Collaboration}\\
        UCL\\
        E-mail: \email{p.litchfield@ucl.ac.uk}}

\abstract{The AlCap experiment is a joint project between the COMET and Mu2e collaborations. Both experiments intend to look for the lepton-flavour violating conversion $\mu + A \rightarrow e + A$, using tertiary muons from high-power pulsed proton beams. In these experiments the products of ordinary muon capture in the muon stopping target are an important concern, both in terms of hit rates in tracking detectors and radiation damage to equipment.  The goal of the AlCap experiment is to provide precision measurements of the products of nuclear capture on Aluminium, which is the favoured target material for both COMET and Mu2e. The results will be used for optimising the design of both conversion experiments, and as input to their simulations. Data was taken in December 2013 and is currently being analysed.}

\FullConference{16th International Workshop on Neutrino Factories and Future Neutrino Beam Facilities - NUFACT2014,\\
		25 -30 August, 2014\\
		University of Glasgow, United Kingdom }

\begin{document}
\section{Muon to electron conversion and the motivation for AlCap}
The term `muon to electron conversion' refers to processes that cause the neutrinoless decay of a muon into an electron, specifically those in which the muon is the ground-state orbit of an atomic nucleus\footnote{Thus, if not specified, it is usually safe to assume we are dealing with negative leptons,  $\mu^-$ and $e^-$. The physics of interest may well be valid for $\mu^+$ as well, but it can only be probed by experiments using free muons.}$\!\!\!\!$.  In this case the conservation of momentum and energy can be achieved by coherent interaction on the nucleus, i.e. $\mu + A \rightarrow e + A$, producing no additional final state particles.  Because of this the electron produced is essentially mono-energetic at around 105\unit{MeV}, this gives distinct signature that can be searched for experimentally.  Other related processes are possible for free and bound muons, namely $\mu \rightarrow e\gamma$ and $\mu \rightarrow 3e$, and experiments to look for these processes are discussed by other speakers \cite{meg, mu3e}.

In the Standard Model all three processes are possible, and at leading order are mediated by a loop containing a $W$ boson and a neutrino. The loop element can be related to neutrino oscillations, which are now well established, so the processes are definitely `in' the Standard Model. However it is strongly suppressed by the mass disparity between the neutrino and $W$ boson, and the expected rate is many orders of magnitude below the sensitivity of any currently conceivable experiment.  So in practice any signal of lepton flavour violation in muon decays is an indication of new physics.

With current accelerator and detector technologies the \mec channel is expected to be the most accessible, and three experiments are in development to study this channel.  The first, DeeMe, uses a low energy primary beam and an innovative combined production and stopping target.\,\cite{deeme}  The other two experiments, COMET \cite{comet} and Mu2e \cite{mu2e}, are designed to take advantage of new high-intensity pulsed muon beams under construction at J-PARC and Fermilab respectively.  Both COMET and Mu2e will (at least initially) use aluminium foils as the stopping target for muons to capture on.    This is broadly speaking the best choice for both experiments, however certain details relating to the aluminium nucleus are not well measured, if at all.  In simulations, these currently have to be extrapolated from data involving similar-mass nuclei for which there is more data, primarily silicon~\cite{Sobottka:1968}.  This is unlikely to be \emph{wildly} wrong, but such extrapolations are rarely perfect and therefore an experiment such as AlCap is desirable.

\subsection* {Why aluminium?}
\begin{figure}
\centering
\includegraphics[width=0.9\textwidth]{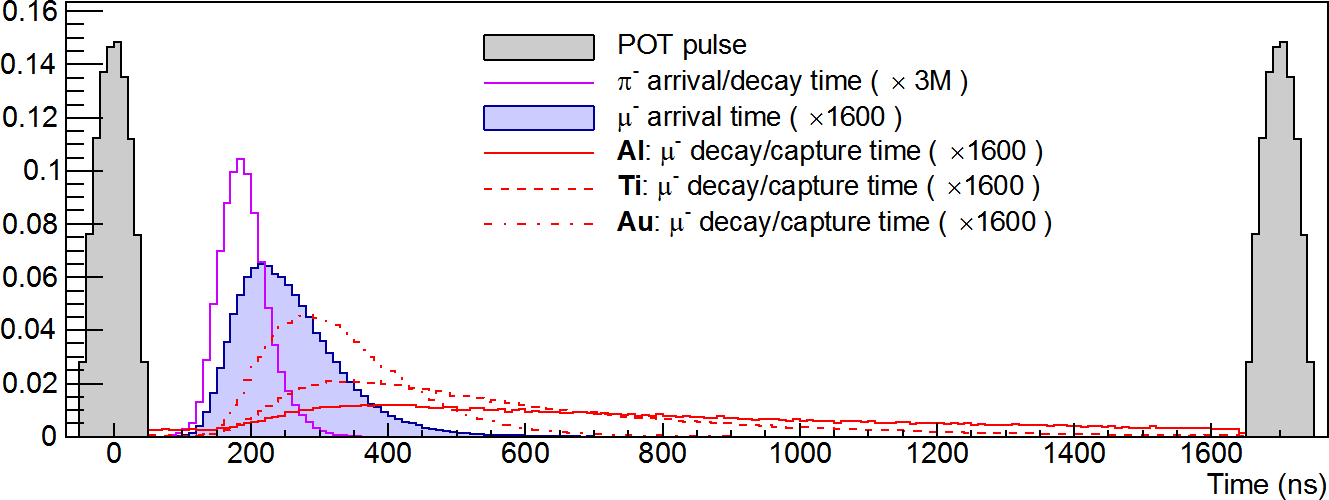}
\caption{Illustration of how $\mu^-$ decay times in various materials would affect Mu2e. The beam pulse occurs roughly once every 1.6\unit{\textmu s}, and prompt backgrounds last for several hundred nanoseconds.  The stopping target material is chosen to maximise the number of captures in the window between beam pulses.  The expected decay curves for aluminium, titanium, and gold are shown. (Figure from Mu2e.)}
\label{mu2e_beam}
\end{figure}

The reason why both experiments wish to use aluminium as a capture target for muons is illustrated in Figure~\ref{mu2e_beam}, which shows the time structure expected for Mu2e.  The bunch spacing at J-PARC will be similar, both a bit more than a microsecond.  Secondary particles can arrive over a period of several hundred nanoseconds, and the detectors will only readout in a delayed window after this pulse has passed. The target material must be chosen so that the decay time for captured muons is matched to this readout window, in order to maximise the number of decays observed.  This decay period, in combination with material requirements leaves aluminium as the first choice, with titanium as an alternative.  The most recent \mec experiment, SINDRUM-II~\cite{Bertl:2006}, had a continuous beam and used a gold capture target because  the number of captures relative to decay-in-orbit increases with nuclear mass.  As can be seen, the lifetime on gold is too short and so it is less suitable for the new generation of experiments.  Aluminium is the preferred target material and is the main focus of the AlCap (`Aluminium Capture') experiment. Because titanium is also suitable, it may be studied in a future run.

\section{Goals of the AlCap experiment}
Because the conversion process is certainly rare, COMET and Mu2e must be designed to cope with a much larger background from regular Standard Model processes. When a muon is captured on an atom it very quickly cascades down to the $1s$ orbital releasing muonic X-rays.  From the $1s$ orbital, the two most important background processes are decay-in-orbit and nuclear muon capture:
\begin{description}
\item[Decay in orbit] is the the process $\mu + A \rightarrow e + \nu\overline{\nu} + A$, which is essentially the same as the decay of a free muon, but with the nucleus able to recoil against the electron.  The nuclear recoil allows a small probability for the electron to carry much more energy than if the muon decayed at rest, up to the 105\unit{MeV} of \mec. This is the main analysis background for COMET and Mu2e.
\item[Nuclear muon capture] (NMC) is the process $\mu + p \rightarrow \nu + n$. This results in an excited nucleus, which will subsequently decay.  Charged decay products are not really a problem for analysis, as typically they are easy to distinguish from the signal, but may result in a large number of tracks in the detectors, reducing resolution and increasing the volume of data that must be recorded and analysed.  The experiment designs must therefore take them into account for the positioning of sensor elements.  Neutral decay products, especially neutrons, can penetrate materials surrounding the target and present a radiation hazard, both to users and hardware.  This must again be reflected in the design, by providing adequate shielding and ensuring components are sufficiently resistant to radiation damage.
\end{description}

As indicated before, the broad goal of the AlCap experiment is to study the capture process from the initial cascade to the subsequent nuclear decays, in order to optimise the design of both experiments and simulation of backgrounds.  In more detail, AlCap is separated into three work packages:
\begin{description}
\item[WP1 -- Charged particle emission:] Measure the total rate and spectrum of charged particles after NMC. We are most interested in protons, which will dominate single-hit rates in the COMET and Mu2e tracking chambers.
\item[WP2 -- Gamma and X-ray emission:]  Measure the rate of X-ray emission, in particular the muonic $K_\alpha$ transition ($2p\rightarrow1s$). This provides the normalisation for rate measurements, and also serves to verify the experimental approach for use in COMET and Mu2e.
\item[WP3 -- Neutron emission:]  Characterise this as well as possible, in order to determine the shielding required, and the appropriate exposure to use for radiation hardness tests.   
\end{description}

The first run of AlCap was planned for 2013, and focuseed on WP1 and WP2, with a later second run intended that would extend these studies based on what we learn from the data, and collect data for WP3.  Currently, discussions are ongoing about when a second run could take place, in parallel with analysis of the 2013 run.

\section{2013 run at PSI} 
A data run was taken in December 2013 at the Paul Scherrer Institute in Switzerland.  This used the $\pi$E1 beam line which has been newly expanded and used for the MuSun Experiment~\cite{musun}. The $\pi$E1 beamline is a DC cloud muon beam, with a spot diameter below 2cm.  By using an electrostatic separator, electron contamination is reduced 10\%.  For AlCap the beam energy was tuned close to 28\unit{$\textrm{MeV}/c$}, with a nominal momentum bite of 1\%.  Naturally, the majority of data was taken in negative polarity but some data was taken in positive polarity for background studies.  

\begin{figure}
\centering
\includegraphics[width=0.7\textwidth]{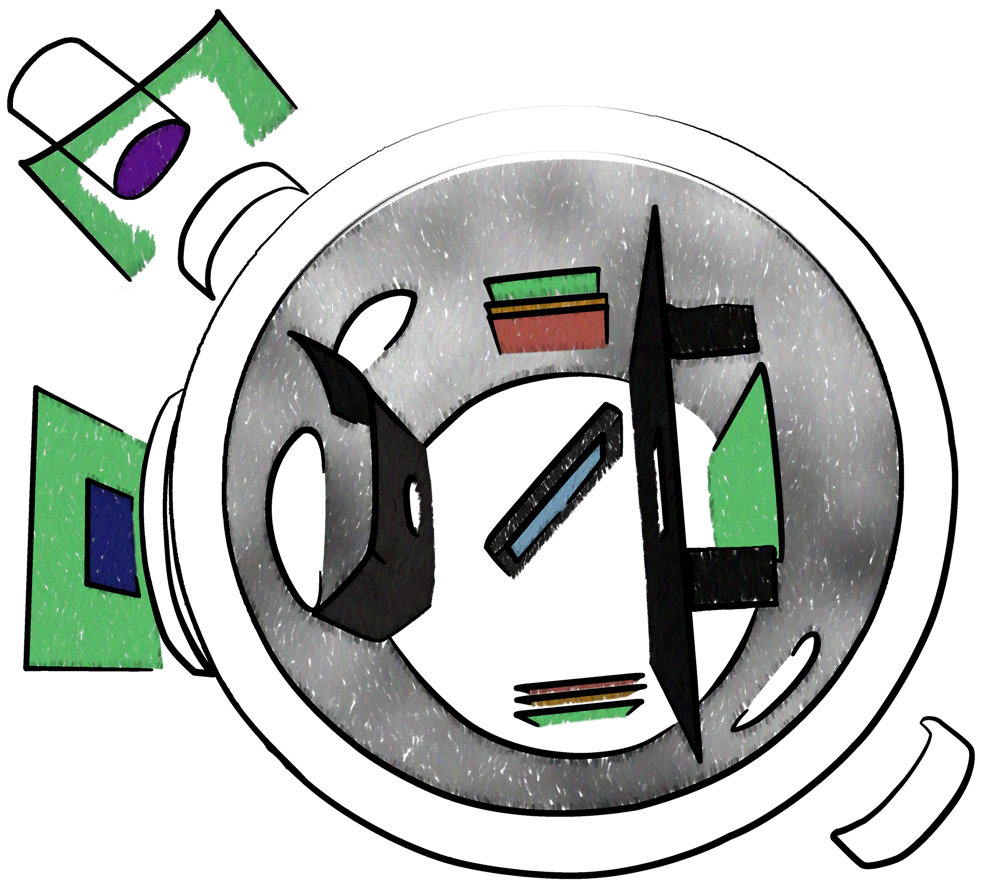}
\caption{Standard configuration during the 2013 run of AlCap. The beam passes through a multiwire position monitor (dark blue) and enters the chamber from the left. It is tuned to stop in the target at the centre (pale blue).  Above and below are shown two `arms' of detector packages (red and orange are silicon) that record particles emitted after muon capture. In the top left corner is a germanium detector (purple) for counting muonic X-rays.  Lead shielding (dark grey) helps to eliminate backgrounds induced by scattered muons, and scintillator paddles (green) are used to count rates in various locations.}
\label{layout}
\end{figure}

The AlCap experiment itself is centred on a cylindrical vacuum chamber roughly 30 centimetres in diameter by 40 centimetres high, with the main access by opening the top plate.   A schematic cartoon of the inside of the chamber seem from above is shown in Figure~\ref{layout}.  The beam enters (from the left on the cartoon) through a Mylar window of standard PSI design. 

A pump is attached to one of the sideports is used to reduce the pressure inside the vacuum chamber to below $10^{-4}\unit{mbar}$. Another sideport provides an aperture for an external germanium detector, and various other ports are used for cable feed-throughs.

\subsection{Beam tuning}
The beam momentum is tuned in 1\% increments in order to maximise the number of muons stopping in the target, as estimated online from count rates in various scintillator hodoscopes.  The first of these of these hodoscopes is used to tag the arrival of charged particles in the beam.  Adjacent to this is a wire chamber that monitors the beam profile in horizontal and vertical planes.  Beam information is used to identify muon candidates in the analysis, and to veto events where a second beam particle is observed.  At the back of the chamber a second hodoscope is used to veto electrons and during the beam tuning.  At momenta around 30\unit{$\textrm{MeV}/c$} electrons are more penetrating than muons, so a low rate of coincident counts is expected from beam contamination.  If the momentum is tuned too high, muons start to punch through the target and the rate in the veto scintillator rises sharply.  After the scan is completed, the beam momentum is tuned based on the results and longer runs are taken.

\subsection{Elements inside the vacuum chamber}
The target sits in the beam, at the centre of the vacuum chamber.  The detector arms sit on either side of the target, along the horizontal line perpendicular to the beam.  This maximises their distance from the beam at a given distance from the target, and therefore minimises backgrounds from scattered beam.  The target oriented at $45^\circ\!\!\!$, so that the left (right) side arm views the front (back) face of the target. This arrangement means it is not so important to know how deep in the target material individual muon stopped, which would otherwise add an uncertain energy loss as the decay products exit the target.

The targets themselves are in the form of thin foils, held in place by a frame-shaped mount. Because the frame sits in the beam halo it is covered with lead foil. Other lead collimators are placed before and after the target. This shielding ensures that, in the region where the detectors are, the beam muons can only capture on the target material or the lead foils.  Light nuclei exposed to beam would produce capture decay products that are a background to the aluminium we wish to study. Negative muons have a much shorter lifetime in lead though, so the background arising from lead elements can be eliminated by imposing a minimum delay requirement between the beam muon tag and the decay products. 

Various targets were used during the run.  Two aluminium foils of different thickness  (100\unit{\textmu m} and 50\unit{\textmu m}) were studied; by having two thickneses we can cross check the energy loss of particles in the foil. In addition, silicon detectors were used as an active target, to \textit{a}) verify the analysis with an extra information; and \textit{b}) cross-check against previous results on silicon \cite{Sobottka:1968}.  A 65\unit{\textmu m}-thick segmented silicon detector was used, but readout problems meant that it was mostly used passively, so other runs were taken using the silicon of the right detector arm as a target. 

In each arm three detector arms are assembled into a single package.  Closest to the target is a thin `transmission' silicon detector which has $2\times2$ segmentation. This detector is 65\unit{\textmu m} thick and used to measure the $\frac{\delta E}{\delta x}$ of incident particles. The next is a much thicker (1.5\unit{mm}) silicon detector that provides enough depth to stop protons from the decay, and hence measures the overall energy.  Finally there is a scintillator plane which is used to tag more penetrating particles.  The silicon detectors were read out over two parallel channels: a slow shaped channel with good energy resolution, and a fast unshaped channel used for accurate timing.

\subsection{External detector}
Detectors for neutral particles can be placed outside the vacuum chamber, since the experiment is small enough that neutral particles can escape with little attenuation.  During the run we used three: a high-purity germanium detector for X-rays produced by the muon dropping down to the atomic ground state; and two neutron detectors.  The neutron detectors were used in a small number of runs to gain experience and some data to develop ideas for the a later run, but were not the focus of this run period.  

The high-purity germanium detector (plus an associated scintillator paddle) was positioned in alignment with one of the side ports. This puts the detector at a greater distance from the target, but this is more than compensated by the thinner port window, in comparison to the vessel walls.  The line we are most interested in is the $2p\rightarrow1s\ (\mu)$ line, which is observed in a known fraction of atomic muon captures, allows us to normalise the rate of other processes.  Corrections for geometric factors and scattering can be evaluated using a Monte Carlo simulation, which itself can be verified by comparing it to data using an active silicon target.  Because the X-ray cascade happens very quickly, a short coincidence window near a tagged beam muon eliminates almost all background, and the resolution of the detector means all the lines of interest are well resolved. 



\section{Analysis}
\begin{figure}
\begin{minipage}[c]{0.4\columnwidth}
\centering
\includegraphics[width=0.95\columnwidth]{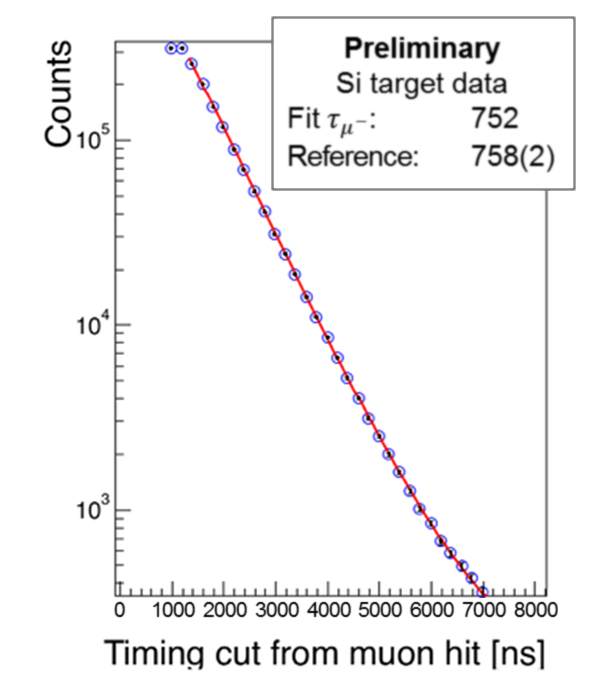}
\end{minipage}
\begin{minipage}[c]{0.6\columnwidth}
\includegraphics[width=0.95\columnwidth]{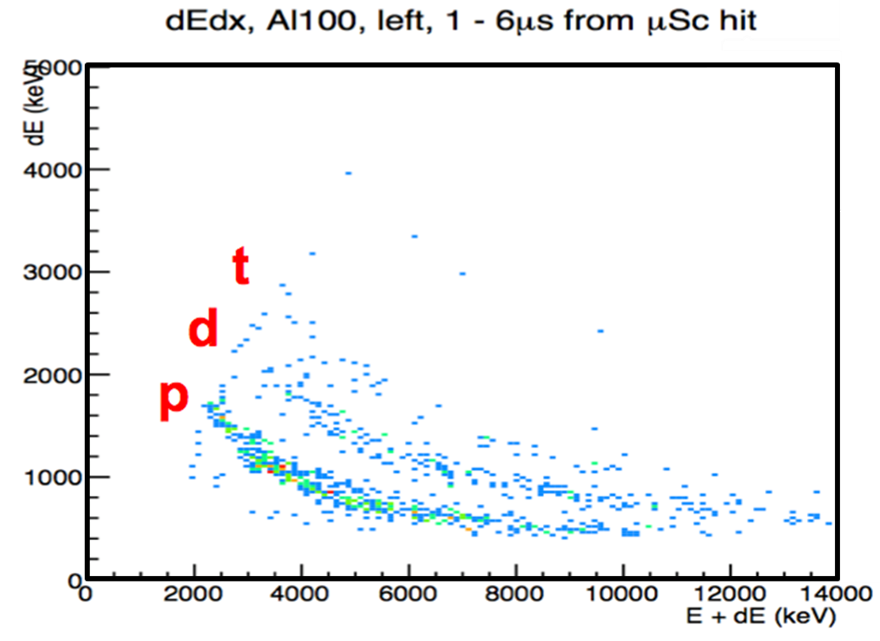}
\end{minipage}
\caption{\textbf{(Left)} Timing of hits in detector arms with respect to the matched muon beam tag.  A fit to an exponential decay including background terms results in a measured lifetime of the $\mu^-$ in silicon very close to the published result, confirming that the majority of events selected come from decays of captured muons. \textbf{(Right)} A plot of the energy in the thin silicon ($\delta E$) as a function of the total energy deposited in both thick and thin silicon ($E + \delta E$), showing the proton / deuteron / triton region.  This simple $\mathrm{d}\ \!\!\!E\! /\! \mathrm{d}\ \!\!\!x$ measurement nevertherless provids an excelent PID. }
\label{ana_plots}
\end{figure}

The analysis of the data is presently ongoing, with a basic chain in place.  Refinements, such as in the algorithms to combine fast and slow pulses, are being developed and will lead to improvements, but the analysis is complete enough to see that the expected features are present.  Because there are many improvements still to come, only preliminary versions of basic plots are shown here, in Fig~\ref{ana_plots}. The left hand plot shows the timing of signals in the Si target after a low-level selection.  The measured time constant indicates that the majority of events after 1000\unit{ns} come from the decay of captured muons.  The right hand plot shows the energy loss in the thin silicon ($\delta E$) as a function of the total particle energy in both thick and thin silicon ($E + \delta E$).  The clear separation between bands demonstrates how these measurements can be used for particle identification, with the lowest is identified as being due to protons.  The clarity of the bands also suggests this data can be used to cross check the energy scale of the detectors---providing a supplement to dedicated calibration runs taken using radioisotopes.  

Although the experiment is designed to require very little input from Monte Carlo simulations, a GEANT4-based simulation does exist, and can be used to refine the analysis and account for systematic uncertainties.  Low-level data, such as the plots shown in Fig~\ref{ana_plots} are being used to tune and validate this Monte Carlo.  Importantly, there is a wealth of information that is separate from the properties we actually wish to measure. For instance, the shape and rate information that are the ultimate goal of WP1 correspond to the colour axis on the the right-hand plot, but the particle bands are sufficiently clear that their shape and location can be considered separately from the colour axis. 

\section{Another run?}
Although the analysis of the 2013 run is not yet complete, there appears to be sufficient data after quality cuts that analyses of the charged particle spectrum and normalisation using X-rays (WP1 \& WP2) are possible using just this run. However there are several further studies we would like to do.  Neutron studies (WP3), and data on a titanium target were in the original plan for a second run.  Additionally, the it would be good to get the segmented silicon detector working and revisit the measurements on a thin active target, as at present we have only a small quantity of data in this configuration.   We are currently working on a request for more beam time at PSI in 2015, and it would be beneficial to do this sooner, before to many students and postdocs move on.

\section*{Acknowledegments}
The AlCap collaboration is a joint effort by groups within the COMET and Mu2e collaborations, which are hosted at J-PARC and FNAL respectively. COMET institutions participating in AlCap are: Osaka University in Japan; IHEP Bejing in China; and Imperial College London and UCL in the United Kingdom.  Mu2e institutions participating in AlCap are: Argonne National Lab, Boston University, Brookhaven National Lab, University of Houston, and University of Washington, all in the United States.  Both COMET and Mu2e collaborations kindly provided figures to be used in this presentation. The speaker is a member of the COMET collaboration.

\end{document}